\documentclass[superscriptaddress,showpacs,floats,twocolumn,prb,floatfix]{revtex4}
\usepackage{amsmath}
\usepackage{amssymb}
\usepackage{bm}
\usepackage{graphicx,epsfig,color}

\begin{document}

\title{Quest for Order in Chaos: Hidden Repulsive Level Statistics\\
       in Disordered Quantum Nanoaggregates}

\author{R.\ Augulis}
\thanks{Present address: Chemistry Center, Lund University,
Getingevägen 60, S-22241, Lund, Sweeden}
\affiliation{Zernike Institute
for Advanced Materials, University of Groningen, Nijenborgh 4, 9747
AG Groningen, The Netherlands}

\author{A.\ V.\ Malyshev}
\thanks{On leave from Ioffe Physiko-Technical Institute, 26
Politechnicheskaya str., 194021 St.-Petersburg, Russia}
\affiliation{Zernike Institute for Advanced Materials, University of
Groningen, Nijenborgh 4, 9747 AG Groningen, The Netherlands}
\affiliation{GISC, Departamento de F\'{\i}sica de Materiales,
Universidad Complutense, E-28040 Madrid, Spain}

\author{V.\ A.\ Malyshev}
\affiliation{Zernike Institute for Advanced Materials, University of
Groningen, Nijenborgh 4, 9747 AG Groningen, The Netherlands}

\author{A. Pug\v{z}lys}
\thanks{Present address: Photonics Institute, Vienna University of
Technology, Gusshausstrasse 27/387, 1040, Vienna, Austria}
\affiliation{Zernike Institute for Advanced Materials, University
of Groningen, Nijenborgh 4, 9747 AG Groningen, The Netherlands}

\author{P.\ H.\ M.\ van Loosdrecht}
\affiliation{Zernike Institute
for Advanced Materials, University of Groningen, Nijenborgh 4, 9747
AG Groningen, The Netherlands}

\author{J.\ Knoester}
\affiliation{Zernike Institute
for Advanced Materials, University of Groningen, Nijenborgh 4, 9747
AG Groningen, The Netherlands}


\begin{abstract}

The local distribution of exciton levels in disordered
cyanine-dye-based molecular nano-aggregates has been elucidated
using fluorescence line narrowing spectroscopy. The observation of a
Wigner-Dyson-type level spacing distribution provides direct
evidence of the existence of level repulsion of strongly overlapping
states in the molecular wires, which is important for the
understanding of the level statistics, and therefore the `functional
properties, of a large variety of nano-confined systems.

\end{abstract}


\maketitle

One of the current dreams in the
field of molecular optics is the full understanding of nature's way
to harvest and use photonic energy which ultimately could enable the
development and design of highly efficient functional optical
devices using molecular arrangements as building blocks. One of the
crucial elements of such photonic assemblies are the 'wires' which
transport the energy between the different functional units of the
devices. Natural systems often utilize structures of coupled
aggregated pigments to transport energy in the form of excitonic
excitations.~\cite{vanAmerongen00,Renger01,Berlin06,Scholes06} Such
structures can also be mimicked in synthetic systems, greatly
assisting studies aiming to understand their fundamental properties.
An important class of synthetic species, on which we focus here, is
found in the so-called one dimensional (1D) $J$-aggregates based on,
for instance, pseudoisocyanine, porphyrin, and benzimidazole
carbocyanine dyes.~\cite{Kobayashi96,Knoester02,Pugzlys06}

Synthetic, as well as natural systems, usually exhibit a substantial
degree of disorder, arising from the environment and from vibrations
and disorder within the systems themselves. In general the presence
of disorder in gapped systems leads to the formation of highly
localized states inside the optical, electronic or magnetic energy
gap of the unperturbed system;  {\it i.e.} to a tail of the density
of states inside the gap generally referred to as the Lifshits
tail.~\cite{Lifshits88} There are many systems which optical
properties are governed by exciton-like excitations highly
susceptible to disorder leading to localization and level repulsion
phenomena. These include conjugated oligomer
aggregates~\cite{Spano06} and polymers,~\cite{Hadzii99} molecular
$J$-aggregates,~\cite{Kobayashi96,Knoester02,Pugzlys06}
semiconductor quantum wells and quantum dots,~\cite{Takagahara03}
gold nanoparticles, \cite{Kuemmeth08} semiconductor quantum
wires,~\cite{Akiyama98} as well as photosynthetic light harvesting
complexes~\cite{vanAmerongen00,Renger01} and
proteins~\cite{Berlin06} (see Ref.~\onlinecite{Scholes06} for a
recent overview). In all these systems, excitons are confined at
least in one dimension at a nanometer scale.

The physical and transport properties of most of the above mentioned
systems are predominantly determined by the states residing in the
vicinity of the energy gap, {\em i.e.} the gap excitation itself and
the Lifshits tail below it, even at finite
temperatures.~\cite{Malyshev03} The localization of the exciton
states within the Lifshits tail gives rise to a local (hidden)
statistics of the levels, which deviate substantially from the
overall statistics.~\cite{Malyshev95,Malyshev01a} Therefore,
understanding the physical properties of these systems requires the
use of statistical approaches; the energy level distributions become
an important part of the theory and interpretation of the
experimental data.

For non-interacting systems it is well
known that the presence of disorder leads to an energy spectrum with
a Poissonian energy spacing distribution. In less trivial cases  of
interacting systems, the situation naturally becomes more complex
leading to the concept of level repulsion, {\it i.e.} a vanishing
probability to find two quantum states with the same energy. The
level statistics of such a system is known as Wigner-Dyson
statistics (see the excellent textbook by Metha
Ref.~\onlinecite{Metha04} for an overview).

Level repulsion phenomena in nano-confined materials has recently
drawn considerable attention, in particular, concerning localized
Wannier excitons in disordered quantum
wells~\cite{Haacke01,Intonti01} and wires,~\cite{Feltrin03} and in
disordered graphene quantum dots,~\cite{Libisch09} as well as
concerning vibronic states in polyatomic
molecules.~\cite{Krivohuz08} Time-resolved resonant Rayleigh
scattering~\cite{Haacke01} and near-field
spectroscopy~\cite{Intonti01,Feltrin03} have been used to study
them. In Ref.~\onlinecite{Malyshev07}, an alternative method has
been proposed to analyze  the level statistics -- low-temperature
time-resolved selectively excited exciton fluorescence spectroscopy,
widely known as fluorescence line narrowing (FLN) spectoscopy. Under
a narrow (compared to the the $J$-band width) excitation within the
$J$-band, the fluorescence spectrum consist of a sharp intensive
peak at the excitation energy and, growing in time, a red-shifted
feature, resulting from the exciton band relaxation. It is this
feature that contains information about the level statistics of the
spatially overlapped states.

Here we apply a variant of this method to reveal the repulsive
statistics of levels residing in the Lifshits tail~\cite{Lifshits88}
of J-aggregates of pseudoisocyanine (PIC) with a chloride
counter-ion (PIC-Cl). In contrast to the earlier proposal to use
time dependent FLN,~\cite{Malyshev07} we here show that also steady
state FLN, which is much more simple in realization, can be utilized
to extract the desired information. We extract the conditional
probability distribution of the repelled states from the
experimental data and show that this distribution is
Wigner-Dyson-like indicating zero probability for zero energy
spacing, which is a fingerprint of the level repulsion.

\begin{figure}[h]
\centerline{\includegraphics[width=\columnwidth,clip]{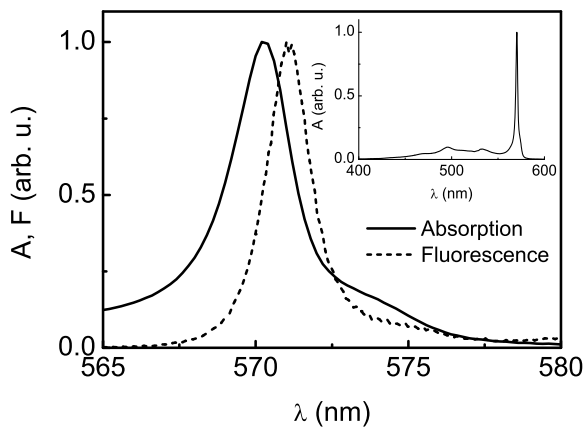}}
\caption{
Low temperature steady-state absorption (solid line) and  fluorescence
(dotted line) spectra of $J$-aggregate of PIC-Cl in the neighborhood
of the $J$-band. The fluorescence spectrum was measured after
off-resonance excitation far in the blue tail at temperature $T = 4$
K. The inset show the absorption spectrum in a wider spectral range.
} \label{Non selective spectra}
\end{figure}

Figure~\ref{Non selective spectra} shows the absorption $A$ and
fluorescence $F$ spectra of $J$-aggregates of PIC-Cl, the latter
measured after excitation using 400 nm light. Both spectra exhibit
an intense peak arising from the dominant exciton transitions (the
$J$-band) and a much less intense and broad shoulder located on the
red side of the $J$-band. We relate this red feature to aggregates
in the vicinity of the substrate as the relative intensity of this
shoulder decreases upon increasing thickness of the aggregated film.
For completeness, we note that the overall absorption spectrum of
our samples  (see inset in Fig.~\ref{Non selective spectra}) is
found to be in good agreement with earlier results.~\cite{Renge97}

\begin{figure}[ht]
\centerline{\includegraphics[width=\columnwidth,clip]{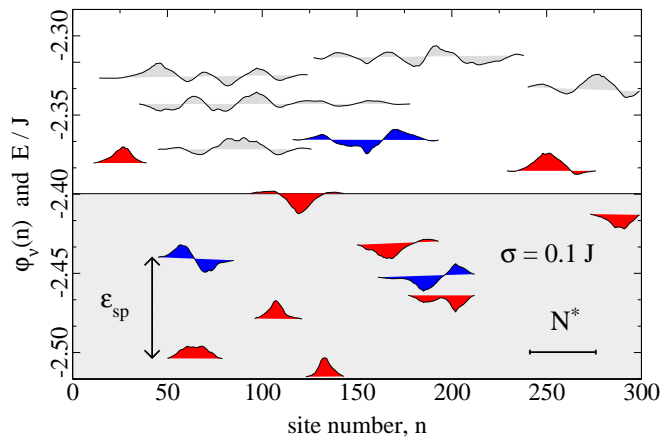}}
\caption{A typical realization of the exciton wave functions
$\varphi_{\nu n}$ ($\nu =1\ldots 14$) in the neighborhood of the
bare exciton band edge $E_b/J = -2.404$. The Lifshits tail of the
DOS ($E < E_b$) is shaded. The origin of the energy is chosen at
$E_0 = 0$, the baseline of each wavefunction represents its energy
in units of $J$. Wave functions are in arbitrary units. $N^*$
denotes the typical localization size of the tail states. Filled red
curves are $s$-like states which overlap weakly. Some of them appear
even slightly above the bare band edge. Filled blue curves are
$p$-like states which overlap well with their $s$-like partner state
lying below. Higher grey-shaded states are band states. They are
delocalized to a larger extent as compared to the tail states. }
\label{FirstStates-color}
\end{figure}

Before turning to the main experimental results, we briefly sketch 
some of the theoretical background relevant for the present study; 
more details on the model and the energy level structure of the 
disordered $J$-aggregates has been discussed in Ref.~\onlinecite{Malyshev01a}. 
A typical realization of the calculated low-energy level structure 
and wavefunctions for a one dimensional aggregate of 300 chromophors 
is depicted in Fig.~\ref{FirstStates-color}. In this calculation, we used 
a gaussian disorder distribution of the chromophore energies with 
standard deviation $\sigma=0.1 J$ (disorder degree from  now on), 
$J$ being the transfer interaction between chromophores (for more 
details see Ref.~\onlinecite{Malyshev01a}).

Without disorder wavefunctions are fully delocalized, and the
lowest state is located at -2.404$J$. The presence of disorder leads
to localization of the wavefunctions within so called segments, and
to the appearance of highly localized states within the band gap,
{\em i.e.} in the Lifshits tail (grey shaded area in Fig.~\ref{FirstStates-color}). 
These are the states of our primary interest since they determine 
the optical properties and transport in the J-aggregates.
They originate from localization in well-like fluctuations of the 
site potential on the molecules. The optically dominant states 
resemble $s$-like wavefunctions which have no nodes within their 
localization segments. The $s$-like states lying deep in the Lifshits 
tail usually appear as singlets and are localized by the so called 
optimal fluctuations of the site energy.~\cite{Lifshits88} Close to 
the band edge, however, the $s$-like states often have partners 
localized within the same localization segment. The latter look like 
$p$-states, having one node within their localization segment. 
Manifolds like these form the local (hidden) structure of the tail 
of the density of states. Since these states are localized on the 
same segment, one may expect level repulsion to occur for them, as 
is indeed observed. In contrast, states from distant (non-overlapping) 
manifolds can be arbitrarily close in energy.

Optical experiments probe the states with a finite transition dipole
moment. For the $s$-like states, the transition dipole moment scales
proportionally to $\sqrt{N^\ast}$, where $N^\ast$ is the typical
localization length of the states. This enhancement of the dipole
moment is known as superradiant enhancement.~\cite{Fidder90} Typically,
the $p$-like states have a transition dipole moment which is several
times smaller than the $s$-like states.~\cite{Malyshev07}
Nevertheless, since the $p$-states are not perfectly antisymmetric,
they do have a finite transition dipole moment and these states can
be optically excited too. Therefore optical experiments can be used
to probe the level statistics by studying the relaxation between $p$
and $s$ like levels.~\cite{Malyshev07}

\begin{figure}[ht]
\centerline{\includegraphics[width=\columnwidth,clip]{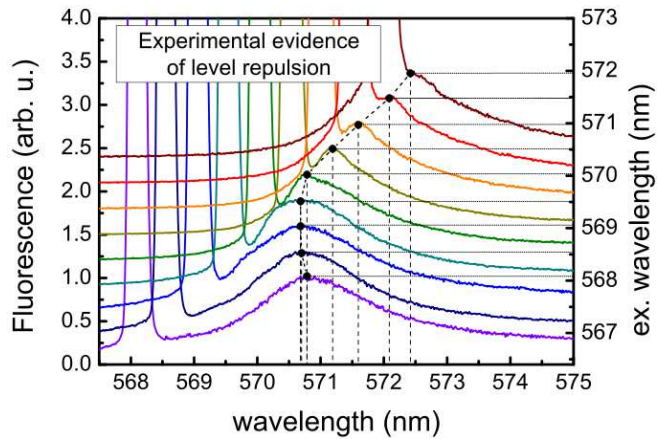}}
\caption{
Low-temperature ($T = 4$ K) fluorescence spectra of $J$-aggregates
of PIC-Cl measured while selectively exciting within the $J$-band.
The vertical lines show the positions of the red feature maxima. The
excitation wavelengths are indicated along the right vertical axis.
} \label{FLN spectra}
\end{figure}

To experimentally study the level statistics in the neighborhood of 
the exciton band edge, we performed steady state resonance fluorescence 
measurements using a narrow excitation line to excite states within 
the $J$-band. Figure~\ref{FLN spectra} shows a number of such spectra 
recorded at low temperature using different excitation energies. The 
spectra show a strong peak at the excitation wavelength, together 
with a broad red-shifted emission band separated from the main peak 
by a pronounced dip. The red shifted emission originates from 
relaxation of the initially excited exciton states into states of 
the Lifshits tail of the DOS. The position of the maximum of this 
band remains almost unchanged while $J$-aggregates are excited on 
the blue side of the $J$-band. For the red-side excitation, the peak 
position moves to the red, the line shape changes considerably, and 
the dip washes out. This is a consequence of the changes in the
relaxation pathways since red-side excitation predominantly excites
$s$-like states.

The dip close to the excitation energy in blue side excited spectra
distribution shows that energy relaxation into states close to the
excitation energy is substantially suppressed, hinting to the 
occurrence of level repulsion. The sheer existence of the dip, 
however, is not enough to conclude on the level statistics. The 
problem is that the relaxation process from the initially excited 
states to the lower lying levels is phonon assisted and, hence, the 
line shape is determined by the product of the level spacing 
distribution function and the phonon spectral density. Since the 
phonon spectral density vanishes for zero energy, one expects the 
spectral intensity to vanish at the excitation energy, even without 
level repulsion. Moreover, one also should bear in mind that together 
with the fluorescence of the relaxed excitons two more processes 
contribute to the red feature and affect its line shape: the phonon 
side-band fluorescence,~\cite{Malyshev07} and surface-mediated 
fluorescence in our thin samples. All three contributions to the 
red-shifted feature are spectrally superposed and must be separated 
in order to extract the signal we are interested in. We note, 
however, that if the observed red-shifted feature would solely 
originate from the phonon side-band, its line shape and position 
would be virtually independent of the excitation wavelength, clearly 
in contradiction to the experimental observations.

In the analysis of the observed spectra we limit ourselves to those spectra
measured using excitation on the blue side of the $J$-band, since it is here
that one expects the $p$-like states to contribute most strongly. In order
to discriminate the true relaxation mediated fluorescence (RMF) from the
surface mediated and the phonon sideband fluorescence we use a simple
subtraction method. For this, we consider the differential spectrum
between two experimental spectra with close excitation wavelengths
$\lambda_2 > \lambda_1$ defined by
\begin{equation}
\label{Differential spectrum}
\Delta F(\lambda_1,\lambda_2,\lambda)
= F(\lambda_{2},\lambda) - \beta\,F(\lambda_{1},\lambda-\lambda_{2}
+ \lambda_{1})\ ,
\end{equation}
where the second term on the right hand side is the $F(\lambda_1,\lambda)$
spectrum shifted in wavelength to match its excitation peak position with that
of the
$F(\lambda_{2},\lambda)$ spectrum. In addition, this term is rescaled by
a factor of $\beta$ in order to cancel the red tail in the spectra; any feature
that is not wavelength dependent is suppressed in the difference spectrum
\ref{Differential spectrum}, and the resulting difference spectrum
represents just the RMF differential signal $\Delta R(\lambda_1,\lambda_2,\lambda)$:
\begin{equation}
    \label{After subtraction}
\Delta R(\lambda_1,\lambda_2,\lambda) = R(\lambda_{2},\lambda) -
\beta\,R(\lambda_{1},\lambda)\ ,
\end{equation}

At the next step, we calculated the quantum efficiency of the
red-shifted feature and found that it did not exceed 0.3 for most
blue excitation. For the spectra we will use in the fitting
procedure, the efficiency even smaller, around 0.1. This means that
excitons make only one step of relaxation, moreover, the major
contribution to this process comes from the intra-segment hops (see
the discussion in Section 2). Then the theoretical RMF line shape
$R(\lambda_e,\lambda) \sim S(\lambda - \lambda_e)
P_{sp}(\lambda_e,\lambda - \lambda_e)$, and we can relate two RMF
spectra taken for different (close) excitation wave length
$\lambda_{2} > \lambda_{1}$ as:
\begin{equation}
R(\lambda_{2},\lambda) \approx
\frac{S(\lambda-\lambda_{2})}{S(\lambda-\lambda_{1})}\,
R(\lambda_{1},\lambda) \ , \label{spectra_relation}
\end{equation}
where we assumed that the energy spacing distribution function
varies much slower than the phonon spectral density, the assumption
which, as will be seen, is consistent with the final results.
Substituting Eq.~(\ref{spectra_relation}) into (\ref{After
subtraction}), we arrive at a relationship between the differential
and ordinary RMF spectra:
\begin{equation}
\Delta R(\lambda_1,\lambda_{2},\lambda) \approx g(\lambda)
R(\lambda_{2},\lambda) \ ,
\label{difference_spectrum}
\end{equation}
where
\begin{equation}
g(\lambda)=
1-\beta\,\frac{S(\lambda-\lambda_{1})}{S(\lambda-\lambda_{2})} \ ,
\label{correction_function}
\end{equation}
i.e., the lineshape of the RMF spectrum can be extracted from the
lineshape of the differential RMF spectrum by dividing it by the
known correction function, provided the factor $\beta$ is adjusted
to cancel long red tale.

It is important that the correction function is almost constant for
wave lengths that are far from the excitation wave lengths
($\lambda-\lambda_{1,2}\gg|\lambda_{2}-\lambda_{1}|$). Hence, it
would not change the shape of the distant features in the long red
tail of the experimental fluorescence spectra. Such features will
therefore be canceled in the difference spectrum which will contain
only the contribution of the RMF. Finally, applying the above
reasoning and formulae to the experimental FLN spectra, we can
recover the MRF lineshape $R(\lambda_{e},\lambda)$ according to the
formula (\ref{difference_spectrum}).
%
%
\begin{figure}[ht]
\centerline{\includegraphics[width=\columnwidth,clip]{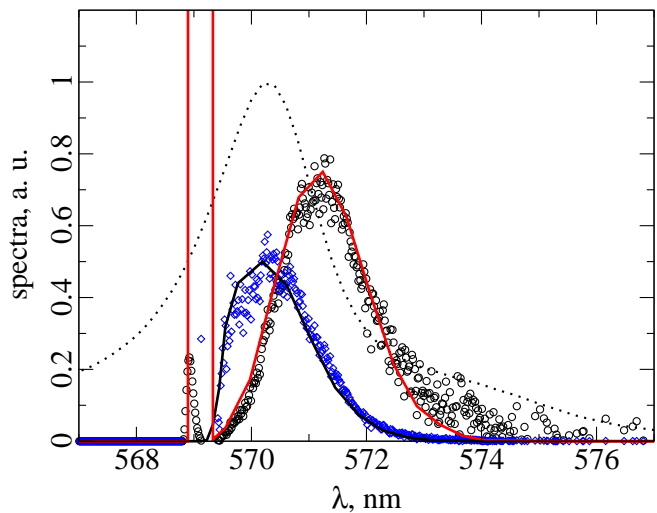}}
\caption{
The conditional probability of the nearest level spacing
distribution $P_{sp}$ (diamonds) obtained by dividing the
experimental curve (circles) by the Debye-like spectral density
$S(\lambda) \propto \lambda^{-3}$ together with the calculated level
spacing distribution $P_{sp}$ (dashed curve). The experimental RMF
spectrum (circles) show the data after applying the subtraction
method (described in the text) to eliminate the contribution of the
long red tail resulting from non-RMF transitions. Also shown are the
theoretical RMF spectrum calculated for the Debye-like spectral
density $S(\lambda) \propto \lambda^{-3}$ (solid curve) and the
absorption spectrum (dotted curve).
} \label{P12_alpha=3}
\end{figure}

The described procedure has been
applied to two spectra recorded using excitation in the blue part of
the absorption spectrum ($\lambda_1=568.5$~nm and
$\lambda_2=569$~nm). The result, assuming a Debye-like spectral
density $S(\lambda)\propto\lambda^{-3}$ is shown in
Fig.~\ref{P12_alpha=3}. The extracted RMF spectrum, obtained
using $\beta=1.06$, is shown in the figure by the open circles.
Clearly, the non-RMF contributions, leading to the long red tail the
fluorescence spectrum, is, as expected, nearly fully eliminated.
Superimposed on this experimental spectrum is the result of a
simulation of the RMF fluorescence spectrum, using a Gaussian
disorder with standard deviation $\sigma=0.2$, again assuming a
Debye model for the phonon spectral density, and an exciton-phonon
scattering strength of $W_0 =22.4$~J (see for details
Ref.~\onlinecite{Malyshev07}), which reproduces the Stokes shift
presented in Fig.~\ref{Non selective spectra}.

\begin{figure}[ht]
\centerline{\includegraphics[width=\columnwidth,clip]{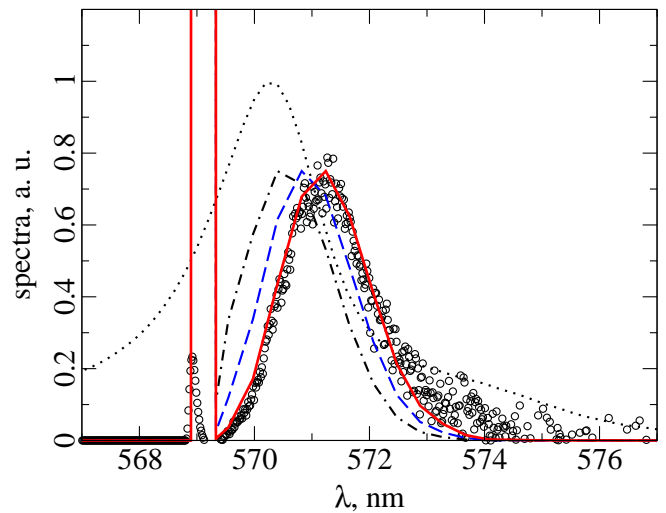}}
\caption{
Open circles represent the experimental red-shifted feature obtained
after applying the subtraction approach (described in the text) to
eliminate a contribution of the long red tail resulting from non-RMF
transitions. The solid curve is the theoretical RMF spectrum
calculated for the Debye spectral density $S(\lambda)\propto
\lambda^{-3}$, while the dashed-dotted and dashed curves are the RMF
spectra for $S(\lambda) \propto \lambda$ and $S(\lambda) \propto
\lambda^{-2}$, respectively. The dotted curve denotes the absorption
spectrum.
} \label{DifferentAlphas}
\end{figure}

It is to be noticed that calculations performed with non-Debye
models for the phonon spectral density, do not lead to a
satisfactory agreement with the experimental data. This is clearly
demonstrated in Fig.~\ref{DifferentAlphas}. The apparent validity of
the usage of the Debye model corroborates the results of
Ref.~\onlinecite{Heijs05}, where this model has been successfully
used to explain the temperature dependence of the J-band width and
the radiative lifetime of J-agregates of the dye pseudoisocynine
wiith different counter ions.

Figure~\ref{P12_alpha=3} shows the principal result of the present
work: the conditional distribution of the nearest-level spacing,
$P_{sp}(\lambda_e,\lambda - \lambda_e)$ (diamonds), obtained after
dividing the extracted RMF spectrum presented in
Fig.~\ref{P12_alpha=3} (open circles) by the phonon spectral density
$S(\lambda) \sim \lambda^{-3}$. We see that
$P_{sp}(\lambda_e,\Delta\lambda)$ tends to zero upon $\Delta\lambda
\to 0$. This is a clear signature of the repulsive statistics of the
nearest level spacing. i.e., it is of a Wigner-Dyson-type. In spite
of the observed distribution is strikingly close to a Wigner-Dyson
one, it remains difficult to determine whether it is the trully
Wigner-Dyson distribution, requiring a linear decrease to zero
probability for zero spacing, which, though consistent with the data,
is not fully proven by the current experiments and analysis.

To conclude, we experimentally studied the statistics of the low 
energy spectrum of disordered molecular nano aggregates of pseudoisocyanine
with the chloride counter ion in the neighborhood of the exciton
band edge. The fluorescence line narrowing technique, allowing to
probe the local energy level distribution,~\cite{Malyshev07}  has
been exploited for this goal. We found a clear signature of a
Wigner-Dyson-like distribution for  the nearest level spacing,
originating from the exciton states localized on the same segment of
the aggregate and thus undergoing the quantum mechanical level
repulsion. This is the first direct experimental prove of the
existence of hidden structure of the exciton low energy spectrum,
the region which dominates the aggregate optical response and
low-temperature transport.

Finally, we note that our finding has a wider applicability than 
for a simple 1D Frenkel exciton system considered here. The reason 
is that the nature of the band edge states (mostly in the Lifshits 
tail) is shared by a large variety of systems, such as gold 
nanoparticles\cite{Kuemmeth08}, quantum wells and quantum 
wires~\cite{Alessi00,Klochikhin04,Feltrin04a}.

\noindent{\bf Acknowledgments.} This work is part of the research
program of the Stichting voor Fundamenteel Onderzoek der Materie
(FOM), which is financially supported by the Nederlandse Organisatie
voor Wetenschappelijk Onderzoek (NWO). A. V. M. and V. A. M.
acknowledge support from NanoNed, a national nanotechnology
programme coordinated by the Dutch Ministry of Economic Affairs. A.
V. M. also acknowledges support from the program Ram\'on y Cajal
(Ministerio de Ciencia y Tecnolog{\'\i}a de Espa{\~n}a).

\end{document}